\documentclass[12pt,usenatbib]{mn2e}
\usepackage{graphicx}
\usepackage{epsf}
\usepackage{lscape}
\usepackage{longtable}
\usepackage{rotating}
\usepackage{color}

\newcommand{\lesssim}{\la}
\newcommand{\aap}{A\&A}
\newcommand{\araa}{ARA\&A}
\newcommand{\apj}{ApJ}

\newcommand{\mnras}{MNRAS}

\begin{document}

\title[Extragalactic Constraints on the Initial Mass Function]{Extragalactic Constraints on the Initial Mass Function}

\author[Stephen M. Wilkins, Andrew M. Hopkins, Neil Trentham \& Rita Tojeiro]  
{
Stephen M. Wilkins$^{1}$\thanks{E-mail: smw@ast.cam.ac.uk}, Andrew M. Hopkins$^{2}$, Neil Trentham$^{1}$ \& Rita Tojeiro$^{3}$ \\
$^1$ Institute of Astronomy, University of Cambridge, Madingley Road, Cambridge, CB3 0HA, United Kingdom\\ 
$^2$ School of Physics, University of Sydney, NSW 2006, Australia\\
$^3$ Institute for Astronomy, University of Edinburgh, Royal Observatory, Blackford Hill, Edinburgh EH9 3HJ\\
}
\maketitle

\begin{abstract}
{
The local stellar mass density is observed to be significantly lower than
the value obtained from integrating the cosmic star formation history (SFH),
assuming that all the stars formed with a Salpeter initial mass function
(IMF). Even other favoured IMFs, more successful in reconciling the observed
$z=0$ stellar mass density with that inferred from the SFH, have difficulties
in reproducing the stellar mass density observed at higher redshift.
In this study we investigate to what extent this discrepancy can be alleviated
for any universal power-law IMF. We find that an IMF with a high-mass slope
shallower ($2.15$) than the Salpeter slope ($2.35$) reconciles the
observed stellar mass density with the cosmic star formation history,
but only at low redshifts. At higher redshifts $z>0.5$ we find that observed
stellar mass densities are systematically lower than predicted from
the cosmic star formation history, for any universal power-law IMF.
}				   
\end{abstract} 


\section{Introduction}

A number of studies have noted that the observed local stellar mass
density, assuming a universal initial mass function (IMF) equivalent
to the Salpeter (1955) initial mass function (IMF), is significantly
smaller than that inferred from the cosmic star formation history
(SFH). Integration of the SFH of Hopkins \& Beacom (2006; hereafter HB06)
assuming a Salpeter IMF suggests a local stellar mass-density
of $\Omega_{*;SFH}\sim 0.0066 \pm
0.0015$ (in units of the critical density) whereas analysis of large
galaxy surveys suggest a value $\Omega_{*;obs}\sim 0.0041 \pm 0.0010$
(Wilkins, Trentham \& Hopkins 2008 hereafter WTH08). This behaviour
continues at higher redshifts where the observed stellar mass-density
history (SMH) remains systematically lower than that inferred from the
SFH (HB06, WTH08) as shown in Figure 1.

Both the observed SMH and that predicted from the SFH are dependent upon
the assumed IMF, although the scaling for each is different (see WTH08,
also \S\,3 and \S\,4 below). The commonly assumed IMFs of Kroupa (2001) and
Chabrier (2003) result in a marginally better correspondence between the
observed SMH and that predicted from the SFH (as shown in the lower panel of
Figure 1 for the Kroupa 2001 IMF). Note the lower normalisation of the data
points in this panel compared to those assuming the Salpeter IMF, and the fact
that while the Kroupa (2001) IMF scales down the SMH inferred from the SFH, it
also scales down the observed SMH data, although not by as much. While this
IMF choice reduces the $z=0$ discrepancy slightly, there remains a significant
inconsistency at higher redshift. The IMF of Baldry \& Glazebrook
(2003, hereafter BG03), goes further toward resolving this inconsistency
(Hopkins \& Beacom 2008) although even then the higher redshift discrepancy
remains. It is possible that there remains some systematic errors in the
SFH or SMH that may explain the observed discrepany. These include issues such
as uncertainties in the extent of dust obscuration on the SFH,
or redshift-dependent effects associated with observable rest-frame
wavelengths for the SMH. These are discussed by WTH08 who conclude that
significant systematic errors, at the level required to resolve the
discrepancy, seem unlikely.

There have recently been several indications that the IMF may indeed not be
universal, but instead evolves with redshift. These have arisen from
a variety of independent considerations. Van Dokkum (2008), for example,
invokes evolution of the IMF to reconcile the rate of evolution in both
luminosity and colour for early-type cluster galaxies at moderate redshifts
($z<0.8$). Dav{\'e} (2008) finds that an evolving IMF better explains the
observed evolution in the relationship between individual galaxy star
formation rates and stellar masses. Baugh et al.\ (2005) requires an IMF
with a flatter high-mass slope at higher redshifts to explain the observed
numbers of sub-mm galaxies, and several other studies in the past
few years also favour a flatter IMF at higher redshift (see references
in WTH08).

We put aside the issue of an evolving IMF for the purposes of the current
study. Here we are motivated by the different sensitivity in the scaling
of the SMH and the SFH to the choice of IMF, and we extend the analysis
of WTH08 by considering the extent to which any universal power-law IMF
can reconcile the observed SFH and SMH data. In \S\,2 we introduce the stellar,
integrated galaxial and cosmic initial mass functions, and discuss the
possible distinctions. In \S\,3 and \S\,4 we quantify the effect of the
IMF upon both the observed stellar mass density and that inferred from
the SFH. In \S\,5 we use a $\chi^{2}$ minimisation to establish the
universal IMF that best reconciles the observed SMH with the SFH and
in \S\,6 we present our conclusions. Throughout this work, we assume a
flat $\Lambda$ CDM cosmology with $\Omega_{\Lambda}=0.7$, $\Omega_{\rm
matter}=0.3$ and $H_{0} = 70\,\, {\rm kms}^{-1}\, \rm{Mpc^{-1}.}$

\begin{figure}
\includegraphics[width=20pc]{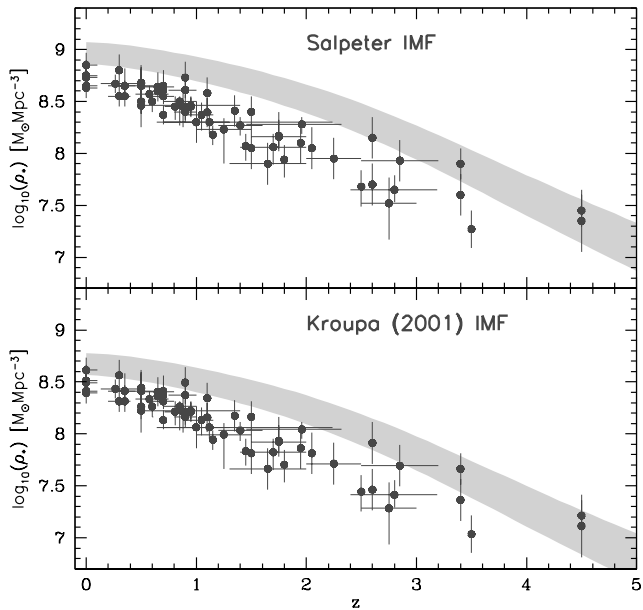} 
\caption{The evolution of stellar mass density assuming a Salpeter IMF (top panel) and a Kroupa (2001) IMF (bottom panel). The shaded region is the stellar mass density inferred from the uncertainty envelope of the SFH of HB06. The points are observed values from compilation of WTH08.}
\end{figure}

\section{The Initial Mass Function}

The mass distribution of stars in recently-formed individual clusters is determined by the stellar initial mass function. The stellar IMF is usually defined such that $dN=\xi(m)\,dm$, where $dN$ is the number of stars in the interval $[m,m+dm]$. An early measurement of the shape of the stellar IMF was conducted by Salpeter (1955) who found it could be well represented within the range $0.4<M/M_{\odot}<10$ by a single power law, i.e. $\xi\propto m^{-\alpha}$, with $\alpha=2.35$ (the Salpeter index). Subsequent studies found that this IMF, if extrapolated, overestimated the abundance of low mass stars ($<0.5\,M_{\odot}$). In general, the stellar IMF is typically found to be well represented by an $n$-part law (although numerous other parameterisations exist), i.e.
\begin{equation}
\xi(m)=k_{i}m^{-\alpha_{i}}\,\,\,m_{i}<m/M_{\odot}<m_{i+1}\,\,\,i\in \{ 0,1,...,n\}
\end{equation}

The stellar IMF is bound by a minimum and maximum allowed stellar mass ($m_{min;*}$ and $m_{max;*}$ respectively). Although studies often vary in their choice of $m_{min;*}$ (extragalactic studies generally adopt $0.1\,M_{\odot}$) the actual minimum stellar mass likely lies around $0.01\,M_{\odot}$. The maximum stellar mass, although still not precisely known, lies in the range $100-200\,M_{\odot}$ (Kroupa 2007). Estimates of both the SMH and SFH are not strongly dependent on the exact choice of the maximum stellar mass, however, because of the steep power-law nature of most IMF estimates. 

The stellar IMF describes the initial mass distribution of stars forming in stellar clusters, but the average IMF over a whole galaxy, referred to as the Integrated Galaxial IMF (IGIMF) may have a rather different form (Kroupa \& Weidner 2003). The distinction between the IGIMF and the stellar IMF arises as a consequence of some fraction of individual stellar clusters being insufficiently massive enough to fully sample the high-mass end of the stellar IMF. This effect causes both small clusters, and consequently the average IMF for the entire galaxy, to be deficient in high-mass stars compared to the stellar IMF. The extent of this effect is influenced by both the distribution of cluster masses and the way small clusters are populated by stars. Kroupa \& Weidner (2003) suggest this effect causes a significant steepening of the IGIMF relative to the stellar IMF. On the other hand Elmegreen (2006) suggests the effect is much smaller, a result obtained from an alternative distribution of cluster masses and procedure to populate small clusters with massive stars. 

A further complication arises from the possibility that the IGIMF is not the same for all galaxies (Weidner \& Kroupa 2005). In such a situation the average IMF of stars formed in all galaxies, the cosmic IMF, may not be equivalent to either the stellar IMF or the IGIMF of a single galaxy. 

The principal consideration of these effects is that inferring the cosmic IMF from either the stellar IMF or the IGIMF is challenging. An alternative method, arising from the global constraints of the SFH and SMH may then be more productive. In this work we consider the effect of an invariant cosmic IMF upon estimates of the stellar mass density inferred both from direct observations and the integration of the SFH. 

For convenience we define a base cosmic IMF $\xi_b$ that we will subsequently allow to be modified. Modifications will be indicated by notation such as $\xi_b(\alpha_2=\alpha_2')$ to denote an IMF identical to the base IMF apart from an alternative high-mass slope $\alpha_2 = \alpha_2'$. We choose this base IMF to be consistent with local studies of the stellar IMF (summarised in Kroupa 2007), which indicate that the stellar IMF can be well described by a three-part power law with
\begin{equation}
\begin{array}{ll}
\alpha_{0}=0.3 & m_{0}=0.01\\
\alpha_{1}=1.3 & m_{1}=0.08\\
\alpha_{2}=2.35 & m_{2}=0.5\\
 & m_{3}=150.\\
\end{array}
\end{equation}

\section{The Star Formation History}

Star formation rates (SFRs) of galaxies are principally obtained from emission associated with short-lived massive stars. These stars, whose lifetimes are comparable to star formation timescales, dominate galactic output of short-wavelength (predominantly ultraviolet) radiation. As such, both ultraviolet luminosities (UV) and gas-reprocessed emission (such as H$\alpha$) are good proxies for the presence of ongoing star formation. Alternative indicators include core collapse supernovae (CCSN) rates, infrared and radio luminosities and many others (See, e.g., Kennicutt 1998 and Hopkins 2004 for an overview). 

The relationship between a given indicator luminosity is determined principally by the physical mechanism responsible for its production and the IMF. The relationship between the luminosity (or CCSN rate in the case of CCSN) and the inferred SFR is usually described by a calibration factor, $A_{i}$ (i.e. $SFR=A_{i}\,L_{i}$), which encapsulates the physical mechanism and the effect of an assumed IMF.

Changes to the low-mass ($<0.5\,M_{\odot}$) end of the IMF are accounted for analytically. These stars generally contribute very little to the various SFR indicators, and are accounted for only when extrapolating over the entire IMF. For example, the ratio of the calibration factors for our base IMF (in the mass range $0.01-150\,M_{\odot}$) to that assuming a Salpeter IMF (in the mass range $0.1-100\,M_{\odot}$)\footnote{Extragalactic studies generally use calibrations derived assuming a Salpeter IMF over the range  $0.1-100\,M_{\odot}$} (i.e. $A_{i\xi_{b}}/A_{i\xi_{sal}}$) is given simply by $\int_{0.01}^{150}m\,\xi_b(m)\,dm/\int_{0.1}^{100}m\,\xi_{sal}(m)\,dm\sim 0.74$ assuming that the high-mass slopes are similarly normalised.

Changes to the IMF outside the low-mass range on the other hand produce a somewhat more complicated effect upon the various calibration factors. In general changes to the IMF which produce a larger fraction of high-mass stars (such as flattening the high-mass slope) naturally reduce the SFR calibrations. This is complicated by the fact that different indicators are driven by different stellar mass ranges. The effect of changing the high-mass end of the IMF will thus in general not produce a uniform conversion for all indicators. Because changes to the high-mass slope of the IMF affect very high mass stars ($>20\,M_{\odot}$) more than lower mass stars, $H\alpha$ luminosities (which are strongly dependent on very massive stars) are more sensitive to the slope than the rate of CCSN or the UV luminosity. 

An additional complication arises because the UV calibration is dependent upon the star formation burst length. The extent of this effect is also somewhat dependent on the IMF, albeit to only a small ($\pm 10\%$) degree for a wide range of burst lengths and slopes. We determine the effect on the SFR calibrations of changes to both $\alpha_2$ and $\alpha_1$ by using the PEGASE.2 population synthesis code (Fioc \& Volmerange 1997). These are displayed in both Table 1, where a selection of IMFs are considered, and in Figure 2.

\begin{figure}
\includegraphics[width=20pc]{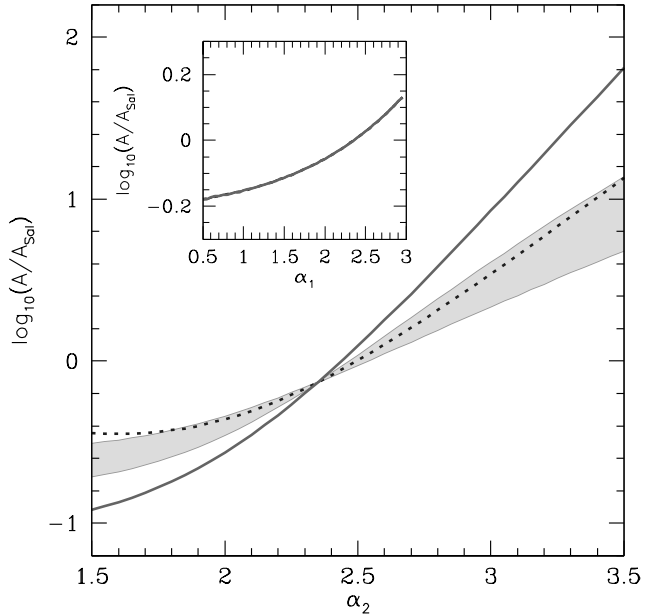} 
\caption{The SFR calibration $A_{i\xi}$ relative to the value for the Salpeter IMF assuming an IMF identical to $\xi_b$ but with an alternative high-mass slope $\alpha_2$. This is shown for H$\alpha$ (solid line), UV (shaded area, indicating uncertainty induced by the effects of different burst lengths) and CCSN rates (dotted line). The inset shows the H$\alpha$, UV and CCSN rate calibrations relative to the value for the Salpeter IMF assuming an IMF identical to $\xi_b$ but with an alternative slope for $\alpha_1$, the slope of the IMF between $0.08\,M_{\odot}$ and $0.5\,M_{\odot}$.}
\end{figure}

\begin{table}
\begin{tabular}{l r r r }
IMF & $H\alpha^{a}$ & $L_{UV}$ & CCSN$^{b}$ \\
\hline
Salpeter$^{c}$ & 1.00 & 1.00 & 1.00 \\
$\xi_{b}$ & 0.74 & 0.74 & 0.74 \\
$\xi_{b}(m_{3}=100\,M_{\odot})$ & 0.85 & 0.76 & 0.73 \\
$\xi_{b}(\alpha_{2}=1.90)$  & 0.22  & 0.37 & 0.39\\
$\xi_{b}(\alpha_{2}=2.15)$ & 0.40 & 0.52 & 0.53 \\
$\xi_{b}(\alpha_{2}=2.70)$  & 2.60 & 1.47 & 1.60 \\
\end{tabular}
\caption{The star formation rate calibrations $A_{i\xi}$ relative to the Salpeter value for a number of different indicators assuming various IMFs. $^{a}$ Here we adopt a calibration based on the average of bursts between $50$ and $1000\,{\rm Myr}$ long. $^{b}$ The core collapse supernovae rate (CCSN) is defined as the number of stars created in range $8<m/M_{\odot}<50$ relative to the total mass of stars created. $^{c}$ Extragalactic star formation rate density and stellar mass density measurements often assume a Salpeter IMF in the range $0.1\le M/M_{\odot}\le 100$, for ease of comparison we quote the values for this IMF, but for the other IMFs we employ the full mass range ($0.01\le M/M_{\odot}\le 150$) discussed in section \S\,2.}  
\end{table}

The stellar mass density implied by the star formation history $\rho_{*;\rm SFH}$ is the integral over the (previous) star formation history corrected for the effects of mass loss through stellar evolution processes:
\begin{equation}
{\rm \rho_{*;SFH}}(\xi,t)=\int_{0}^{t}{\rm SFR_{obs}}(\xi,t')(1-f_{\rm r}[\xi,t-t']) dt',
\end{equation}
where the quantity $f_{{\rm r}}[\xi,t-t']$ encapsulates the processes responsible for returning stellar material to the interstellar medium (such as spernovae and stellar winds). This is the fraction of stellar mass created at $t'$ that has been returned to the ISM by $t$. Because these mechanisms are dependent upon the initial stellar mass, this quantity is dependent on the IMF. In general for an IMF with a larger proportion of high-mass stars (such as an IMF with a flatter high-mass slope) the fraction of material returned to the interstellar medium is larger.  

The result of changing the IMF on $\rho_{*;\rm SFH}$ is then due to two effects, changes to the amount of material returned to the ISM and modification of the SFR calibrations. The combined effect on the relative (to the Salpeter IMF) stellar mass density is shown as a function of the high-mass slope in Figure 3 based on the UV calibration. The effect of changing the slope in the range $0.08<M/M_{\odot}<0.5$ (i.e. $\alpha_{1}$) is shown in the inset to this figure. 

\section{The Observed Stellar Mass Density}

The observed stellar mass density $\rho_{*;\rm obs}$ can be estimated in a number of ways. In the simplest approach, a luminosity density (obtained by integrating a galaxy luminosity function) can be converted into a stellar mass density assuming a mass-to-light ratio (MLR). In this case rest-frame near infrared (NIR) light is preferential because, unlike shorter wavelengths, it is not dominated by young stars but is instead more representative of the entire underlying stellar population. The specific MLR is typically that which would be expected for average stellar population.

An alternative approach relies on determining, and integrating the galaxy stellar mass function. This requires that individual galaxy stellar masses are measured. These again can be estimated using NIR luminosities, however the use of an average MLR is inappropriate because individual galaxy star formation histories are typically diverse. Instead the individual galaxy MLR, which is primarily affected by a galaxy's recent star formation history, can be constrained through the use of additional photometry. For example UV luminosities can be used to determine, and thus correct for, the effect of recent star formation activity.

In the absence of rest-frame NIR observations (prevalent at higher redshift), techniques used for constraining the MLR can be used to infer the stellar mass content alone, albeit with larger uncertainties (see Dye 2008). Most implementations of this technique attempt to match galaxy spectra or broadband luminosities to a library of template spectral energy distributions (SEDs). These template spectra are generated using population synthesis models and generally encompass a range of star formation histories, metallicity distributions as well as dust obscuration corrections. Alternative implementations (such as the VESPA algorithm; Tojeiro et al. 2007) bypass the use of pre-constrained SFHs and age-metallicity relationships. Instead, they model a galaxy as a discrete set of stellar populations of different ages, with each stellar population being free to take any SFR and metallicity value. 

Because of the diversity of methods employed for estimating the stellar mass density, exacerbated by the different spectral ranges probed by each method, the dependence on the IMF from measurement to measurement may not be uniform. We consider the conversions that are obtained from the use of the rest-frame NIR luminosity density and those from SED fitting techniques. In the former case the conversion can be simply calculated by comparing the MLR of a stellar population that formed with a SFH similar to the cosmic SFH for each IMF, achieved using the PEGASE.2 population synthesis code. To determine the effect on masses recovered using SED fitting technique we analyse a random sample of 1000 galaxies from the Sloan Digitial Sky Survey data release 5 (Adelman-McCarthy et al. 2007) (which covers a spectral range of $\sim\,3800\AA-9200\AA$) using an implementation of VESPA with SEDs from PEGASE.2. In both cases we find the dependence on both the high and low mass slope of the IMF to be similar ($\pm 5\%$) as shown in Figure 3.

\begin{figure}
\includegraphics[width=20pc]{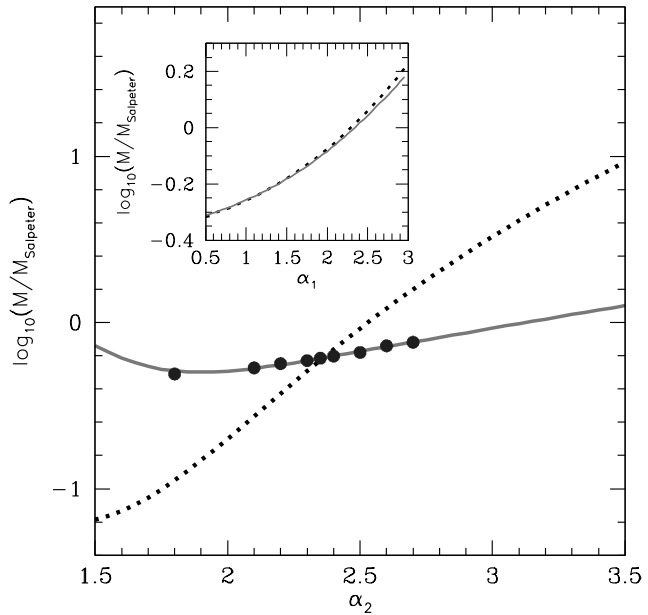} 
\caption{The different dependencies on $\alpha_2$ of $\rho_{\rm *;obs}$ (solid line and points) and of $\rho_{\rm *;SFH}$ (dotted line) as a fraction of that assuming a Salpeter IMF. Solid line: the average NIR MLRs; Points: estimates using VESPA; Dotted line: Masses from integrating SFHs derived from UV luminosities. The inset is similar but shows the dependencies on $\alpha_1$, the slope of the IMF between $0.08\,M_{\odot}$ and $0.5\,M_{\odot}$.}
\end{figure}

\section{Constraining the IMF}

\begin{table}
\begin{tabular}{l r r r r}
IMF  & (1) $\Omega_{*;SFH}$ & (2) $\Omega_{*;obs}$ & (3)\\
\hline
Salpeter$^{a}$  & 0.0066 & 0.0041 & 1.60\\
$\xi_{b}$                        & 0.0040  & 0.0025 & 1.60\\
$\xi_{b}(m_{3}=100\,M_{\odot})$   & 0.0041  & 0.0025 & 1.64\\
$\xi_{b}(\alpha_{2}=1.90)$      & 0.0009  & 0.0020 & 0.45\\
$\xi_{b}(\alpha_{2}=2.15)$      & 0.0021  & 0.0022 & 0.96\\
$\xi_{b}(\alpha_{2}=2.70)$      & 0.0104  & 0.0031 & 3.35\\
\end{tabular}
\caption{The local stellar mass density, in units of the critical density, inferred from the HB06 star formation history (1), and that observed based on an aggregate of studies (see text) for various IMFs. (3) is the ratio of (1) to (2), which is an indication of the disparity between the two densities. $^{a}$ Extragalactic star formation rate density and stellar mass density measurements often assume a Salpeter IMF in the range $0.1\le M/M_{\odot}\le 100$, for ease of comparison we quote the values for this IMF, but for the other IMFs we employ the full mass range ($0.01\le M/M_{\odot}\le 150$) discussed in section \S\,2.}  
\end{table}

Figure 3 highlights the fact that changes to the high mass slope of the cosmic IMF affect both $\rho_{*;\rm obs}$ and $\rho_{*;\rm SFH}$ differently, whereas changes to the low-mass slopes change both equally. Specifically $\rho_{*;\rm SFH}$ is affected significantly more than $\rho_{*;\rm obs}$ by changes to the high-mass slope. A flatter high-mass slope, which in general reduces the inferred stellar mass density, will act to bring $\rho_{*;\rm obs}$ and $\rho_{*;\rm SFH}$ into better agreement. This result is consistent with the flatter high-mass slope found by BG03, $\alpha_2=2.15$, based on fitting population synthesis model SEDs to the local broadband luminosity densities spanning ultraviolet to near-infrared wavelengths.

Exploring a range of high-mass slopes, a $\chi^{2}$ minimisation gives the optimum universal cosmic IMF that brings the local $\rho_{*;\rm obs}$ and $\rho_{*;\rm SFH}$ into best agreement. Using the SFH of HB06 and the local $\rho_{*;\rm obs}$ measurements from the compilation of WTH08 we find this to be $1.95<\alpha_{2}<2.3$, with the best fit $\alpha_2\sim 2.15$, as shown in Figure 4.

The extent of the discrepancy and thus the optimum IMF recovered is due to the precise choice of SFH and local stellar mass density. The WTH08 compilation of stellar mass densities includes many recent studies and while the HB06 SFH similarly includes a broad compilation of measurements, many of these, especially at high redshift, require careful assumptions regarding dust obscuration (see discussion in WTH08). Fardal et al. (2007) similarly determined a SFH based on a compilation of diverse SFR measurements. Due to inclusion of a different set of measurements and different dust assumptions they found a SFH similar to that of HB06 at low redshift but slightly decreased at higher redshifts. This SFH yielded a local stellar mass density around $15\%$ smaller than that of HB06 thus bringing the SFH and local stellar mass density into a closer agreement. Fardal et al. (2007) also considered the correspondence between the extragalactic background light (EBL), the local observed $K$-band luminosity density (a rough proxy for the stellar mass density) and the SFH. They found only poor agreement between the EBL, the SFH and the $K$-band luminosity density when assuming an IMF with a Salpeter high-mass slope. If a BG03 IMF is assumed there is considerably better agreement.

Although this high-mass slope is slightly shallower than the fiducial Salpeter slope it is still marginally consistent with the scatter on measurements of the stellar IMF (Kroupa 2007). However the cosmic IMF is not necessarily equivalent to the stellar IMF. Based on the arguments of Kroupa \& Weidner 2003, though, it is likely that the cosmic IMF is equivalent to, or is steeper than, the stellar IMF. This does suggest there is a significant inconsistency between the estimates of effective IMFs averaged over all local galaxy populations and those measured directly for individual star clusters. 

In recent years a multitude of studies have measured stellar mass densities up to $z\sim 5$ as discussed in WTH08. The top panel of Figure 5 shows $\rho_{*;\rm obs}$ and $\rho_{*;\rm SFH}$ assuming an IMF with a high-mass slope of $2.15$ ($\xi_b(\alpha_2=2.15)$) over the range $0.0<z<5.0$. In the bottom panel logarithmic residuals are displayed. At low redshift ($z<0.5$) this cosmic IMF produces a good correspondence. At higher redshift though, the correspondence is visibily reduced, with residuals increasing to 0.5\,dex (a factor of 3) by $z=1.5$. The stellar mass density predicted from the star formation history at $z>0.5$ is systematically larger than the observed stellar mass density. Using this data and the stellar mass density evolution inferred from the HB06 SFH a minimum $\chi^{2}$ analysis gives the best fitting cosmic IMF. We find that this has a high mass slope of $1.85<\alpha_{2}<2.15$ (with a best fit of $2.00$), somewhat shallower than obtained using low redshift observations alone. Although a cosmic IMF with a slope of $2.00$ produces statistically the best fit to the data, the value of $\rho_{\rm *;SFH}(z=0)$ inferred is significantly smaller than the observed $\rho_{\rm *;obs}(z=0)$. This is an important point, since local estimates of the stellar mass density are expected to be the most robust. They benefit from much larger sample sizes, and because rest-frame NIR measurements are available they are less sensitive than those at high redshift to possible systematic effects in stellar mass estimates. Although the discrepancy at $z=0$ can be eliminated with a flatter IMF, it seems that this solution fails to fully resolve the difference at higher redshifts. This suggests that the shape, and thus the evolution of $\rho_{*;\rm SFH}$ and $\rho_{*;\rm obs}$ is somewhat different. This could arise from a number of systematic effects affecting stellar mass estimates or star formation rates, including for example, uncertainties in the extent of dust obscuration.

Alternatively, WTH08 and a number of other studies (see \S\,1) have suggested that an improved correspondence can be achieved by relinquishing the assumed invariance of the cosmic IMF. Furthermore Weidner, Kroupa, \& Larson (2004) suggest that the mechanism by which the IGIMF is steepened relative to the stellar IMF (see discussion in \S\,2, and Kroupa \& Weidner 2003) is dependent upon the galaxy wide star formation rate. Because of the strong redshift evolution of the star formation rate distribution function, such an effect would give rise to a cosmic IMF which is redshift dependent. In a work in preparation (Wilkins et al in prep) we are investigating a number of different evolutionary scenarios, with respect to the remaining SMH-SFH discrepancy.

\begin{figure}
\includegraphics[width=20pc]{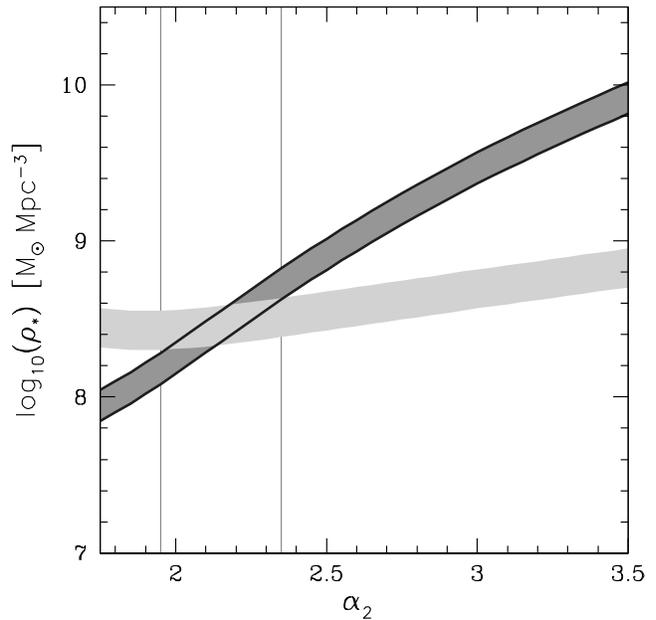} 
\caption{The local (i.e. $z\sim 0$) stellar mass density as predicted from the HB06 star formation history (dark shaded area) and that from observations (light shaded area) assuming an IGIMF equivalent to $\xi_{b}$ but with a high mass slope $\alpha_{2}$. The vertical lines indicate the allowed range of $\alpha_{2}$ ($1.95<\alpha_{2}<2.35$). }
\end{figure}

\begin{figure}
\includegraphics[width=20pc]{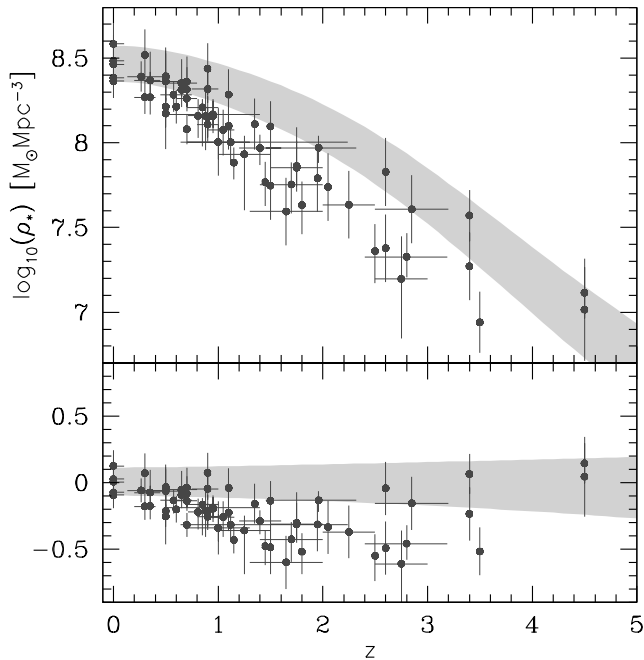} 
\caption{The observed stellar mass density ($\rho_{*;obs}(z)$, points) compared with the stellar mass density implied by the star formation history ($\rho_{*;SFH}(z)$, shaded area) assuming an $\xi_{b}(\alpha_{2}=2.15)$ IMF. The bottom panel shows the logarithmic residual between the $\rho_{*;obs}(z)$ and $\rho_{*;SFH}(z)$.}
\end{figure}

\section{Conclusions}

We have investigated the effect of a universal cosmic initial mass function on the correspondence between the observed build up of stellar mass density (WTH08) and that predicted from the star formation history of HB06. We find that a cosmic IMF with a high-mass slope of $\alpha_{2}=2.15\pm0.15$ can produce a statistical reconciliation between the star formation history and observed stellar mass density at low redshifts ($z<0.5$). At higher redshifts, however, the observed stellar mass density lies systematically below that predicted from the star formation history. The remaining discrepancy, of order a factor of 3 at redshifts $2\lesssim z \lesssim 4$, may be possible to reconcile through systematic errors in the SMH or SFH measurements, although this seems unlikely (see discussion in WTH08). An evolving IMF remains an attractive solution to this problem.

Any possible evolving cosmic IMF must resolve not only the discrepancy between the SMH and SFH, but many other aspects of galaxy evolution (e.g., Dav{\'e} 2008, van Dokkum 2008) together with maintaining consistency with the extragalactic background light (Fardal 2007), and the chemical evolution of galaxies, together with matching the observed local IGIMF. 

\vskip 20pt

\noindent We would like to thank Karl Glazebrook for discussions leading to this paper. We would also like to thank Pavel Kroupa, Carsten Weidner and Thomas Maschberger for useful discussion about the IGIMF. Further, we also thank the anonymous referee for helpful comments and suggestions. SMW acknowledges support of an STFC studentship and of King's College, NT acknowledges support provided by STFC, AMH acknowledges support provided by the Australian Research Council in the form of a QEII Fellowship (DP0557850) and RT is funded by the Funda\c{c}\~{a}o para a Ci\^{e}ncia e a Tecnologia under the reference PRAXIS SFRH/BD/16973/04.

\end{document}